\RequirePackage{fix-cm}
\documentclass[twocolumn,epjc3]{svjour3}
\smartqed  
\RequirePackage{graphicx,color,multirow,array}
\journalname{Eur. Phys. J. D}
\begin{document}
\title{Self-gravitating envelope solitons in a degenerate quantum plasma system}
\author{N. Ahmed\thanksref{e1} \and N.A. Chowdhury \and A. Mannan \and A.A. Mamun}
\thankstext{e1}{e-mail: nahmed93phy@gmail.com}
\institute{Department of Physics, Jahangirnagar University, Savar, Dhaka-1342, Bangladesh}
\date{Received: date / Accepted: date}
\maketitle
\begin{abstract}
  The existence and the basic features of ion-acoustic (IA) envelope solitons in a self-gravitating degenerate
  quantum plasma system (SG-DQPS), containing inertial non-relativistically degenerate light and heavy ion species
  as well as inertialess non-relativistically degenerate positron and electron species, have been theoretically investigated by deriving the nonlinear Schr\"{o}dinger (NLS) equation. The NLS equation, which governs the
  dynamics of the IA waves, has disclosed the modulationally stable and unstable regions for the IA waves. The unstable
  region allows to generate bright envelope solitons which are modulationaly stable. It is found that the
  stability and the growth rate dependent on the plasma parameters (like, mass and number density
  of the plasma species). The implications of our results in astronomical
  compact object (viz. white dwarfs, neutron stars, and black holes, etc.) are briefly discussed.
\end{abstract}
\section{Introduction}
\label{1:Introduction}
The field of self-gravitating degenerate quantum plasma (DQP) physics is one of the
current interesting research field among the plasma physicists because of the painstaking
observational evidence confirms the existence of such kind of extreme plasma in astronomical compact
objects (viz. white dwarfs, neutron stars, and black holes, etc. \cite{Chandrasekhar1931,Koester1990,Fowler1994,Koester2002,Zaman2017})
and potential applications in modern technology (viz. metallic and semiconductor
nano-structures, quantum x-ray free-electron lasers, nano-plasmonic devices \cite{Atwater2007,Stockmann2011},
metallic nano-particles, spintronics \cite{Wolf2001}, thin metal films, nano-tubes, quantum
dots, and quantum well \cite{Manfredi2007}, etc.). The number density of the plasma species is
extremely high in  self-gravitating DQP system (SG-DQPS) (order of $10^{30}~cm^{-3}$ in white dwarfs \cite{Koester1990,Shapiro1983}
and order of $10^{36}~cm^{-3}$ or even more in neutron stars \cite{Koester1990,Shapiro1983}) which
leads to generate a strong gravitational field inside the plasma medium. Basically, the SG-DQPS
contains degenerate inertial light (viz. ${\rm ~^{1}_{1}H}$ \cite{Fletcher2006,Killian2006} or ${\rm ~^{4}_{2}He}$
 \cite{Chandrasekhar1931,Fowler1994} or ${\rm ~^{12}_{~6}C}$ \cite{Koester1990,Koester2002}) and  heavy
(viz. ${\rm ~^{56}_{ 26}Fe}$ \cite{Vanderburg2015} or ${\rm ~^{85}_{ 37}Rb}$ \cite{Witze2014} or ${\rm ~^{96}_{42}Mo}$ \cite{Witze2014})
ion species and inertialess degenerate electron and positron  species.
Heisenberg's uncertainty principle established the relationship between the uncertainty to determine the position
and momentum of a particle simultaneously, and mathematically it can be expressed as, $\Delta x\Delta p\ge \hbar/2$
(where $\Delta x$ is the uncertainty in position of the particle and $\Delta p$ is the uncertainty in momentum
of the same particle, and $\hbar$ is the reduced Planck constant). This indicates that the position of the plasma
species are very certain (because of highly dense and compressed plasma species) inside the plasma system but the
momenta of the plasma species are extremely uncertain. Therefore, these plasma species are uncertain (certain)
in momentum (position) give rise to a very high pressure known as ``degenerate pressure''. The expression for the
degenerate pressure $P_j$ (degenerate plasma particle species $j$) as a function of number density ($N_j$) is given by \cite{Chandrasekhar1931,Fowler1994,Mamun2011}
\begin{eqnarray}
P_j=K_j N_j^{\gamma}, ~~~~ \gamma=\frac{5}{3}, ~~~K_j\simeq\frac{3}{5}\frac{\pi\hbar^2}{m_j},
\label{1eq:1}
\end{eqnarray}
where $j=e$ ($p$) for the electron (positron) species, and $j=l$ ($h$) for the light (heavy)
ion species, respectively. The $\gamma$ is the relativistic factor ($\gamma=5/3$ stands for
non-relativistic case and $\gamma=4/3$ stands for ultra-relativistic case) and $m_j$ is the mass.
It is clear from equation (\ref{1eq:1}) that the degenerate pressure $P_j$ is independent on
thermal temperature but depends on degenerate particle number density $N_j$ and mass $m_j$.
Finally, the strong gravitational field (degenerate pressure) of the SG-DQPS wants to
squeeze (stretch) the plasma system but they are counter-balanced to each other.

During the last few years, a large number of authors have studied the propagation of nonlinear
waves in DQP by considering self-gravitational or without self-gravitational field.
Asaduzzaman \textit{et al.} \cite{Asaduzzaman2017} have investigated the linear and nonlinear
propagation of self-gravitational perturbation mode in a SG-DQPS and found that self-gravitational
perturbation mode becomes unstable when the wavelength of the perturbation mode is minimum.
Mamun \cite{Mamun2017} examined the self-gravito shock structures in a SG-DQPS.
Chowdhury \textit{et al.} \cite{Chowdhury2018} have studied the modulational instability (MI) of
nucleus-acoustic waves in a DQP system and found that the bright and dark
envelope solitons are modulationally stable. But to the best of our knowledge, no attempt
has been made to study MI of the ion-acoustic waves (IAWs) by deriving a nonlinear  Schr\"{o}dinger equation (NLS)
and formation of the envelope solitons in any kind of SG-DQPS. Therefore, in the present
work, a SG-DQPS (containing  inertialess degenerate electron and positron species, inertial
degenerate light as well as heavy ion species) has been considered to obtain the conditions of MI
of the IAWs and the formation of the envelope solitons, and also to identify their basic features.

The manuscript is organized as follows. The basic governing equations for the dynamics of the
SG-DQPS are descried in Sec. \ref{1:Governing equation}. The derivation of the NLS equation
is provided in Sec.  \ref{1:Derivation of NLS equation}. The stability of the IAWs and
envelope solitons are examined in Sec. \ref{1:Stability analysis}. A brief discussion is
finally presented in Sec. \ref{1:Discussion}.
\section{Governing equation}
\label{1:Governing equation}
We consider a SG-DQPS containing inertialess degenerate electrons (mass $m_e$; number density $N_e$),
positrons (mass $m_p$; number density $N_p$), inertial degenerate light ions (mass $m_l$; number density $N_l$),
and heavy ions (mass $m_h$; number density $N_h$). The detail information about the light and heavy nuclei is provided in Table \ref{Table:1}.
The nonlinear dynamics of the SG-DQPS is described by
\begin{eqnarray}
&&\hspace*{0.0cm}\frac{\partial P_e}{\partial X}=-m_e N_e \frac{\partial \tilde{\phi}}{\partial X},
\label{1eq:2}\\
&&\hspace*{0.0cm}\frac{\partial P_p}{\partial X}=m_p N_p \frac{\partial \tilde{\phi}}{\partial X},
\label{1eq:3}\\
&&\hspace*{0.0cm}\frac{\partial N_l}{\partial T} + \frac{\partial}{\partial X} (N_l U_l)=0,
\label{1eq:4}\\
&&\hspace*{0.0cm}\frac{\partial U_l}{\partial T} + U_l\frac{\partial U_l }{\partial X}=-\frac{\partial \tilde{\phi}}{\partial X}-\frac{1}{m_l N_l}\frac{\partial P_l}{\partial X},
\label{1eq:5}\\
&&\hspace*{0.0cm}\frac{\partial^2 \tilde{\phi}}{\partial X^2}=4 \pi G (m_l N_l + m_h N_h + m_e N_e + m_p N_p),
\label{1eq:6}\
\end{eqnarray}
where $P_e$, $P_p$, and $P_l$ are the degenerate pressure of the degenerate electrons,
positrons, and light ions, respectively;
$X$ ($T$) is the space (time) variable; $U_l$ is the light ion fluid speed; $\tilde{\phi}$
is the self-gravitational potential; $G$ is the universal gravitational constant.
Now, the charge neutrality condition for the electrostatic
wave potential is
\begin{eqnarray}
&&\hspace*{0.0cm}N_e=N_p+Z_l N_l +Z_h N_h,
\label{1eq:7}\
\end{eqnarray}
where $Z_l$ and $Z_h$ are the charge state of light and heavy ions, respectively. Here,
it may be noted that the effect of the electrostatic wave potential has been neglected.
Now, we consider normalized variables, namely, $x=X/L_q$, $t=T \omega_{jl}$, $n_l=N_l/n_{l0}$, $u_l=U_l/C_q$,
$C_q=\sqrt{\pi} \hbar n_{e0}^{1/3}/m_l$, $\phi=\tilde{\phi}/C_q^2 $,
$\omega_{jl}^{-1}=(4 \pi G m_l n_{l0})^{-1/2}$ (where $n_{l0}$ and $n_{e0}$ are the
equilibrium number densities of the light ion and electron species, respectively).
After normalization, Eqs. ($2$)-($6$) can be taken in the following form
\begin{eqnarray}
&&\hspace*{0.0cm}\frac{\partial \phi}{\partial x}=-\frac{3}{2} \alpha^2 \frac{\partial n_e^{2/3}}{\partial x},
\label{1eq:8}\\
&&\hspace*{0.0cm}\frac{\partial \phi}{\partial x}=\frac{3}{2} \sigma_1^2 \sigma_2^\frac{2}{3} \frac{\partial n_p^{2/3}}{\partial x},
\label{1eq:9}\\
&&\hspace*{0.0cm}\frac{\partial n_l}{\partial t} + \frac{\partial}{\partial x} (n_l u_l)=0,
\label{1eq:10}\\
&&\hspace*{0.0cm}\frac{\partial u_l}{\partial t} + u_l\frac{\partial u_l }{\partial x}=-\frac{\partial \phi}{\partial x}-\beta \frac{\partial n_l^{2/3}}{\partial x},
\label{1eq:11}\\
&&\hspace*{0.0cm}\frac{\partial^2 \phi}{\partial x^2}= \gamma_e (n_e -1)- \gamma_l (n_l -1) + \gamma_p (n_p -1),
\label{1eq:12}\
\end{eqnarray}
where $\alpha= m_l/m_e $, $\sigma_1 = m_l /m_p $, $ \sigma_2=n_{p0}/n_{e0}$, $\mu=n_{e0}/n_{l0}$, $\beta=(3/2) \mu^{-2/3}$,
$\lambda= n_{p0}/n_{l0}$, $\gamma= Z_l m_h/Z_h m_l$ (which is greater than $1$ for any set of heavy and light ion species), 
$\gamma_e = \mu(1/\alpha + \gamma/Z_l )$ (here, $1/\alpha \ll \gamma/Z_l$, where $1/\alpha$ varies from $\sim 10^{-4}$ to 
$\sim 10^{-3}$, and $\gamma/Z_l$ varies from $\sim 0.1$ to $2.0$, and this means that $\gamma_e \simeq \mu\gamma/Z_l$), 
$\gamma_l=\gamma-1$, $\gamma_p=\lambda(1/\sigma_1 - \gamma/Z_l)$. For inertialess degenerate electron and positron, 
the number densities can be expressed as
\begin{eqnarray}
&&\hspace*{0.0cm}n_e =\left(1-\frac{2 \phi}{3 \alpha^2}\right)^\frac{3}{2},
\label{1eq:13}\\
&&\hspace*{0.0cm}n_p =\left(1+\frac{2 \phi}{3 \sigma_1^2 \sigma_2^\frac{2}{3}}\right)^\frac{3}{2}.
\label{1eq:14}
\end{eqnarray}
Now, we substitute Eqs. (\ref{1eq:13}) and (\ref{1eq:14}) into Eq. (\ref{1eq:12}) and
extend the resulting equation up to third order, we get
\begin{eqnarray}
&&\hspace*{0.0cm}\frac{\partial^2 \phi}{\partial x^2}-\gamma_l +\gamma_l n_l= \gamma_1 \phi + \gamma_2 \phi^2 + \gamma_3 \phi^3 +\ldots,
\label{1eq:15}\
\end{eqnarray}
where
\begin{eqnarray}
&&\hspace*{0.0cm}\gamma_1=\left(\frac{\gamma_p}{\sigma_1^2 \sigma_2^{2/3}}-\frac{\gamma_e}{\alpha^2}\right),\nonumber\\
&&\hspace*{0.0cm}\gamma_2=\left(\frac{\gamma_e}{6 \alpha^4}+\frac{\gamma_p}{6 \sigma_1^4 \sigma_2^{4/3}}\right),\nonumber\\
&&\hspace*{0.0cm}\gamma_3=\left(\frac{\gamma_e}{54 \alpha^6}-\frac{\gamma_p}{54 \sigma_1^6 \sigma_2^2}\right).\nonumber\
\end{eqnarray}
\begin{center}
\begin{table}[h!]
\caption{The values of $\gamma$ when ${\rm ~^{1}_{1}H}$ \cite{Killian2006,Fletcher2006},
${\rm ~^{4}_{2}He}$ \cite{Chandrasekhar1931}, and  ${\rm ~^{12}_{~6}C}$
\cite{Koester1990,Koester2002} are considered as the light ion species,  and ${\rm ~^{56}_{ 26}Fe}$
\cite{Vanderburg2015},  ${\rm ~^{85}_{ 37}Rb}$ \cite{Witze2014}, and  ${\rm ~^{96}_{42}Mo}$ \cite{Witze2014}
are considered as the heavy ion species.}
\begin{tabular}{|p{2.8cm}|p{2.8cm}|m{1cm}|}
\hline
{\bf Light ion species}                                         &{\bf Heavy ion species}                                                             &~~~~${\bf \gamma}$ \\ [2.2ex]
\hline
 \multirow{2}{4em}{~}                                  &~~~~~~${\rm ~^{56}_{ 26}Fe}$  \cite{Vanderburg2015}                                   &~~2.16 \\ [2.2ex]
\cline{2-3}

$~~~~{\rm ~^{1}_{1}H}$ \cite{Killian2006,Fletcher2006}      &~~~~~~${\rm ~^{85}_{ 37}Rb}$  \cite{Witze2014}                                          &~~2.30\\[2.2ex]
\cline{2-3}
 ~                                                      &~~~~~~${\rm ~^{96}_{42}Mo}$  \cite{Witze2014}                                              &~~2.28\\[2.2ex]
\hline
 \multirow{2}{5em}{~}                            &~~~~~~${\rm ~^{56}_{ 26}Fe}$  \cite{Vanderburg2015}                                             &~~1.08 \\[2.2ex]
\cline{2-3}

$~~~~{\rm ~^{4}_{2}He}$ \cite{Chandrasekhar1931}   &~~~~~~${\rm ~^{85}_{ 37}Rb}$  \cite{Witze2014}                                    &~~1.15\\[2.2ex]
\cline{2-3}
 ~                                                                    &~~~~~~${\rm ~^{96}_{42}Mo}$  \cite{Witze2014}                          &~~1.14\\[2.2ex]
\hline
 \multirow{2}{*}{~}                                    &~~~~~~${\rm ~^{56}_{ 26}Fe}$ \cite{Vanderburg2015}                                       &~~1.08 \\[2.2ex]
\cline{2-3}

$~~~~{\rm ~^{12}_{~6}C}$ \cite{Koester1990,Koester2002}   &~~~~~~${\rm ~^{85}_{ 37}Rb}$ \cite{Witze2014}                                        &~~1.15\\[2.2ex]
\cline{2-3}
 ~                                                 &~~~~~~${\rm ~^{96}_{42}Mo}$   \cite{Witze2014}                                             &~~1.14\\[2.2ex]
\hline
\end{tabular}
\label{Table:1}
\end{table}
\end{center}
We note that the terms on the right hand side of Eq. (\ref{1eq:15}) are the contribution of electron and positron species.
Thus, Eqs. (\ref{1eq:10}), (\ref{1eq:11}), and (\ref{1eq:15}) describe the dynamics of the
gravitational envelop solitons in the SG-DQPS under consideration.
\section{Derivation of NLS equation}
\label{1:Derivation of NLS equation}
To investigate the MI of the IA waves in SG-DQPS, we will derive the NLS equation by employing the reductive
perturbation method \cite{Taniuti1969,Chowdhury2017a}. So, we first introduce the stretched co-ordinates for
independent variables $x$ and $t$ in terms of $\xi$ and $\tau$ as follows:
\begin{equation}
\left.
\begin{array}{l}
\xi=\epsilon(x-v_gt),\\
\tau=\epsilon^2t,
\end{array}
\right\}
\label{1eq:16}
\end{equation}
where $v_g$ is the envelope group velocity and $\epsilon$ is
a small dimensionless expansion parameter.
Then we can expand all dependent physical variables $n_l$, $u_l$, and $\phi$ in power series of $\epsilon$ as
\begin{eqnarray}
&&\hspace*{0.0cm}n_l=1+\sum_{m=1}^\infty\epsilon ^{(m)}\sum_{l'=-\infty}^\infty n_{ll'}^{(m)}(\xi,\tau)~\mbox{exp}(il'\Upsilon),
\label{1eq:17}\\
&&\hspace*{0.0cm}u_l=\sum_{m=1}^\infty\epsilon ^{(m)}\sum_{l'=-\infty}^\infty u_{ll'}^{(m)}(\xi,\tau) ~\mbox{exp}(il'\Upsilon),
\label{1eq:18}\\
&&\hspace*{0.0cm}\phi=\sum_{m=1}^\infty\epsilon ^{(m)}\sum_{l'=-\infty}^\infty \phi_{l'}^{(m)}(\xi,\tau)~\mbox{exp}(il'\Upsilon),
\label{1eq:19}\
\end{eqnarray}
where $\Upsilon=kx-wt$ and $k$ ($\omega$) is the real variable representing the fundamental
carrier wave number (frequency). The derivative operators in Eqs. (\ref{1eq:10}), (\ref{1eq:11}),
and ($\ref{1eq:15}$) are regarded as
 \begin{eqnarray}
&&\hspace*{0.0cm}\frac{\partial}{\partial t}\rightarrow \frac{\partial}{\partial t}- \epsilon v_g \frac{\partial}{\partial \xi}+ \epsilon^2 \frac{\partial}{\partial \tau },
\label{1eq:20}\\
&&\hspace*{0.0cm}\frac{\partial}{\partial x}\rightarrow \frac{\partial}{\partial x}+ \epsilon \frac{\partial}{\partial \xi}.
\label{1eq:21}\
\end{eqnarray}
Now, by substituting Eqs. (\ref{1eq:17})-(\ref{1eq:21}) into Eqs. (\ref{1eq:10}), (\ref{1eq:11}),
and(\ref{1eq:15}) and collecting the different powers of $\epsilon$. Now, the first order ($m=1$) reduced
equations with $l'=1$ can be expressed as
\begin{eqnarray}
&&\hspace*{0.0cm}n_{l1}^{(1)}=\frac{k^2}{S} \phi_1^{(1)},
\label{1eq:22}\\
&&\hspace*{0.0cm}u_{l1}^{(1)}=\frac{k \omega}{S} \phi_1^{(1)},
\label{1eq:23}\
\end{eqnarray}
where $S=\omega^2- \beta_1 k^2$ and $\beta_1=2\beta /3$. The compatibility of the system
leads to the linear dispersion relation as
 \begin{eqnarray}
&&\hspace*{0.0cm}\omega^2=\frac{\gamma_l k^2}{\gamma_1+k^2}+ \beta_1 k^2.
\label{1eq:24}
\end{eqnarray}
\begin{figure}[t!]
  \centering
  \includegraphics[width=80mm]{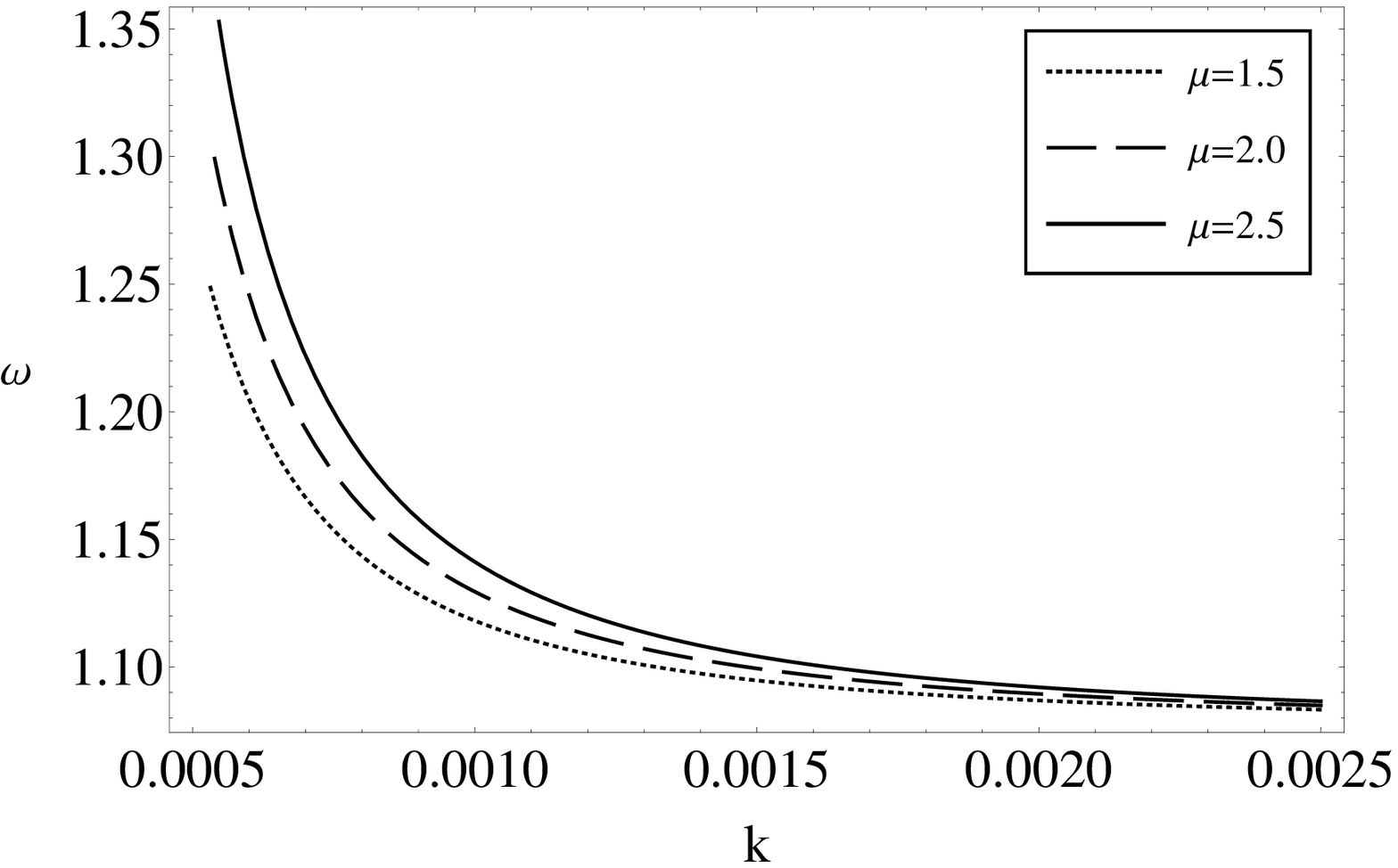}
  \caption{The variation of $\omega$ with $k$ for different values of $\mu$;
  along with $\alpha = 3.67\times10^3$, $\gamma=2.16$, $\gamma/Z_l = 0.5$, $\sigma_1=3.68\times10^3$, $\sigma_2=0.3$, and $\lambda=0.2$.}
  \label{1Fig:F1}
  \vspace{0.8cm}
  \includegraphics[width=80mm]{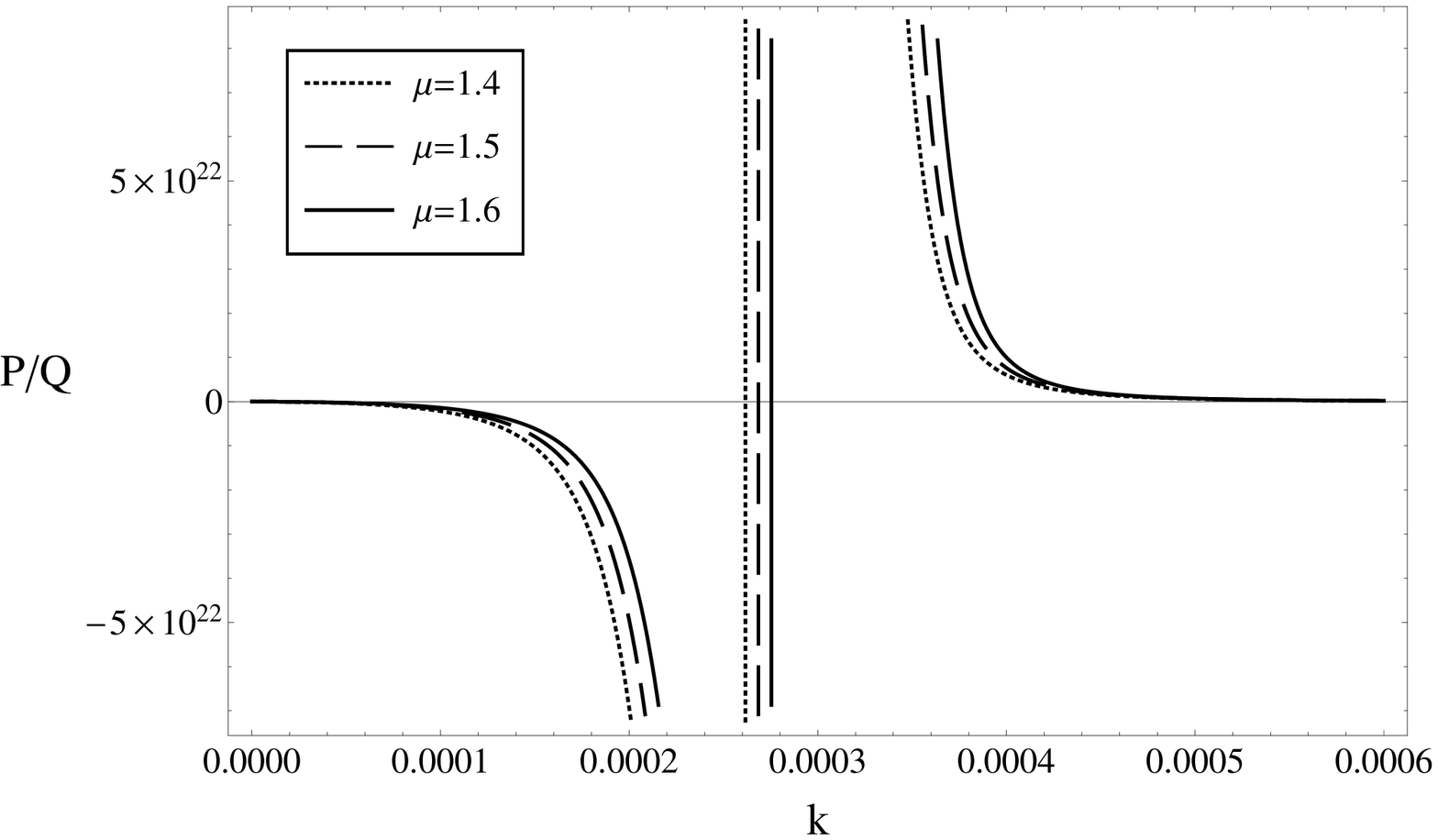}
  \caption{The variation of $P/Q$ with $k$ for different values of $\mu$;
  along with $\alpha = 3.67\times10^3$, $\gamma=2.16$, $\gamma/Z_l = 0.5$, $\sigma_1=3.68\times10^3$, $\sigma_2=0.3$, and $\lambda=0.2$.}
  \label{1Fig:F2}
\end{figure}
The dispersion characteristics of the wave are depicted in Fig. \ref{1Fig:F1} [obtained from
Eq. (\ref{1eq:24})], which indicates that (a) the angular wave frequency ($\omega$) of the IAWs  exponentially decreases
with the increase of $k$; (b) the value of $\omega$ increases with the increase of $n_{e0}$ for the
fixed value of $n_{l0}$ (via $\mu=n_{e0}/n_{l0}$). The second order ($m=2$) reduced
equations with $l'=1$ are given by,
 \begin{eqnarray}
&&\hspace*{0.0cm}n_{l1}^{(2)}=\frac{k^2}{S} \phi_1^{(2)}+\frac{2ik \omega(k v_g - \omega)}{S^2}\frac{\partial \phi_1^{(1)}}{\partial \xi},
\label{1eq:25}\\
&&\hspace*{0.0cm}u_{l1}^{(2)}=\frac{k \omega}{S} \phi_1^{(2)}+\frac{i(\omega^2+ \beta_1 k^2)(k v_g-\omega)}{S^2} \frac{\partial \phi_1^{(1)}}{\partial \xi},
\label{1eq:26}\
\end{eqnarray}
thus, the expression for $v_g$ is obtained as
 \begin{eqnarray}
&&\hspace*{0.0cm}v_g=\frac{\partial \omega}{\partial k}=\frac{\gamma_l \omega^2-(\omega^2 - \beta_1 k^2)^2}{k \omega \gamma_l}.
\label{1eq:27}\
\end{eqnarray}
The amplitude of the second-order harmonics is found
to be proportional to $|\phi_1^{(1)}|^2$ and these are expressed as
\begin{equation}
\left.
\begin{array}{l}
n_{l2}^{(2)} = C_1 |\phi_1^{(1)}|^2,\\
u_{l2}^{(2)} = C_2 |\phi_1^{(1)}|^2,\\
\phi_2^{(2)} = C_3 |\phi_1^{(1)}|^2,\\
n_{l0}^{(2)} = C_4 |\phi_1^{(1)}|^2,\\
u_{l0}^{(2)} = C_5 |\phi_1^{(1)}|^2,\\
\phi_0^{(2)} = C_6 |\phi_1^{(1)}|^2,\\
\end{array}
\right\}
\label{1eq:28}
\end{equation}
where the coefficients are
\begin{eqnarray}
&&\hspace*{0.0cm}C_1=\frac{2 C_3 k^2 S^2 + 3 \omega^2 k^4}{2 S^3},
\nonumber\\
&&\hspace*{0.0cm}C_2=\frac{C_1 \omega S^2 - \omega k^4}{k S^2},
\nonumber\\
&&\hspace*{0.0cm}C_3=\frac{3 \gamma_l \omega^2 k^4 - 2 \gamma_2 S^3}{2 S^3(\gamma_1 +4 k^2)-2 \gamma_l k^2 S^2 },
\nonumber\\
&&\hspace*{0.0cm}C_4=\frac{C_6 S^2 + k^2 \omega^2 + 2 \omega v_g k^3 -\beta_2 k^4}{S^2 (v_g^2-\beta_1)},
\nonumber\\
&&\hspace*{0.0cm}C_5=\frac{C_4 v_g S^2 - 2 \omega k^3}{S^2},~~~~\beta_2 =\beta/9,
\nonumber\\
&&\hspace*{0.0cm}C_6=\frac{(k^2 \omega^2 + 2 \omega v_g k^3 -\beta_2 k^4)\gamma_l - 2 \gamma_2 S^2 (v_g^2-\beta_1)}{\gamma_1 S^2 (v_g^2 -\beta_1)-\gamma_l S^2}.
\nonumber\
\end{eqnarray}
Finally, by substituting all the Eqs. (\ref{1eq:22})$-$(\ref{1eq:28}) into the third order
part ($m=3$) and $l'=1$ and simplifying them, we can obtain the following NLS equation:
\begin{eqnarray}
&&\hspace*{0.0cm}i \frac{\partial \Phi}{\partial \tau} + P \frac{\partial^2 \Phi}{\partial \xi^2}+ Q |\Phi|^2 \Phi=0,
\label{1eq:29}\
\end{eqnarray}
where $\Phi=\phi_1^{(1)}$ for simplicity. The coefficient of dispersion and nonlinear terms $P$ \& $Q$ are given by
\begin{eqnarray}
&&\hspace*{0.0cm}P= \frac{F_1-4 \omega \beta_1^2 k^4-3 k v_g \omega^4 }{2\gamma_l k^2 \omega^2},
\label{1eq:30}\\
&&\hspace*{0.0cm}Q=\frac{S^2[3\gamma_3+2\gamma_2(C_3+C_6)-F_2]}{2\gamma_l \omega k^2},
\label{1eq:31}\
\end{eqnarray}
where
$F_1=4 \beta_1 k^2 \omega^3+2 \beta_1 v_g \omega^2 k^3 + v_g \beta_1^2 k^5$,\\
$F_2=k^2/S^2[2\omega k \gamma_l(C_2+C_5)+\gamma_l \omega^2 (C_1 +C_4)+(\gamma_l \beta_3 k^6/S^2)]$, and $\beta_3=4 \beta/81$.
\section{Stability analysis and envelope solitons}
\label{1:Stability analysis}
Let us now analysis the MI of IAWs by considering the linear solution of the NLS equation (\ref{1eq:29})
in the form $\Phi=\hat{\Phi}~e^{iQ|\hat{\Phi}|^2 \tau}+ c.~c$ (c. c denotes the
complex conjugate), where $\hat{\Phi}=\hat{\Phi}_0+\epsilon \hat{\Phi}_1$ and
$\hat{\Phi}_1=\hat{\Phi}_{1,0}e^{i(\tilde{k}\xi-\tilde{\omega}\tau)}$+ c. c.
Now, by substituting these values into Eq. (\ref{1eq:29}), one readily obtains the
following nonlinear dispersion relation \cite{Chowdhury2018,Sultana2011,Schamel2002,Kourakis2005,Fedele2002,Chowdhury2017b}
\begin{eqnarray}
&&\hspace*{0.0cm}\tilde{\omega}^2=P^2\tilde{k}^2 \left(\tilde{k}^2-\frac{2|\hat{\Phi}_0|^2}{P/Q}\right).
\label{1eq:32}\
\end{eqnarray}
Here, the perturbed wave number $\tilde{k}$ and the perturbed frequency $\tilde{\omega}$ are
different from the carrier wave number $k$ and frequency $\omega$.
It is observed from Eq. (\ref{1eq:32}) that the IAWs will be modulationally stable (unstable) in SG-DQPS
for that range of values of $\tilde{k}$ in which $P/Q$ is negative (positive), i.e., $P/Q<0$ ($P/Q>0$).
When $P/Q\rightarrow\pm\infty$, the corresponding value of $k$ ($=k_c$) is known as the critical or
threshold wave number ($k_c$) for the onset of MI. The variation of $P/Q$ with $k$ for $\mu$ is
shown in Fig. \ref{1Fig:F2} and which clearly indicates that (a) the IAWs are modulatonally
stable (unstable) in SG-DQPS for small (long) wavelength; (b) the $k_c$ increases  with the
increase of $n_{e0}$ for constant value of $n_{l0}$ (via $\mu=n_{e0}/n_{l0}$). In the modulationally unstable ($P/Q>0$) region and under this condition
$\tilde{k}<\tilde{k}_c=\sqrt{2|\hat{\Phi}_0|^2(Q/P)}$, the MI growth rate
can be written [from Eq. (\ref{1eq:32})] as
\begin{eqnarray}
&&\hspace*{0.0cm}\Gamma=|P|\tilde{k}^2 \sqrt{\frac{\tilde{k}^2_c}{\tilde{k}^2}-1}.
\label{1eq:33}\
\end{eqnarray}
\begin{figure}[t!]
  \centering
  \includegraphics[width=80mm]{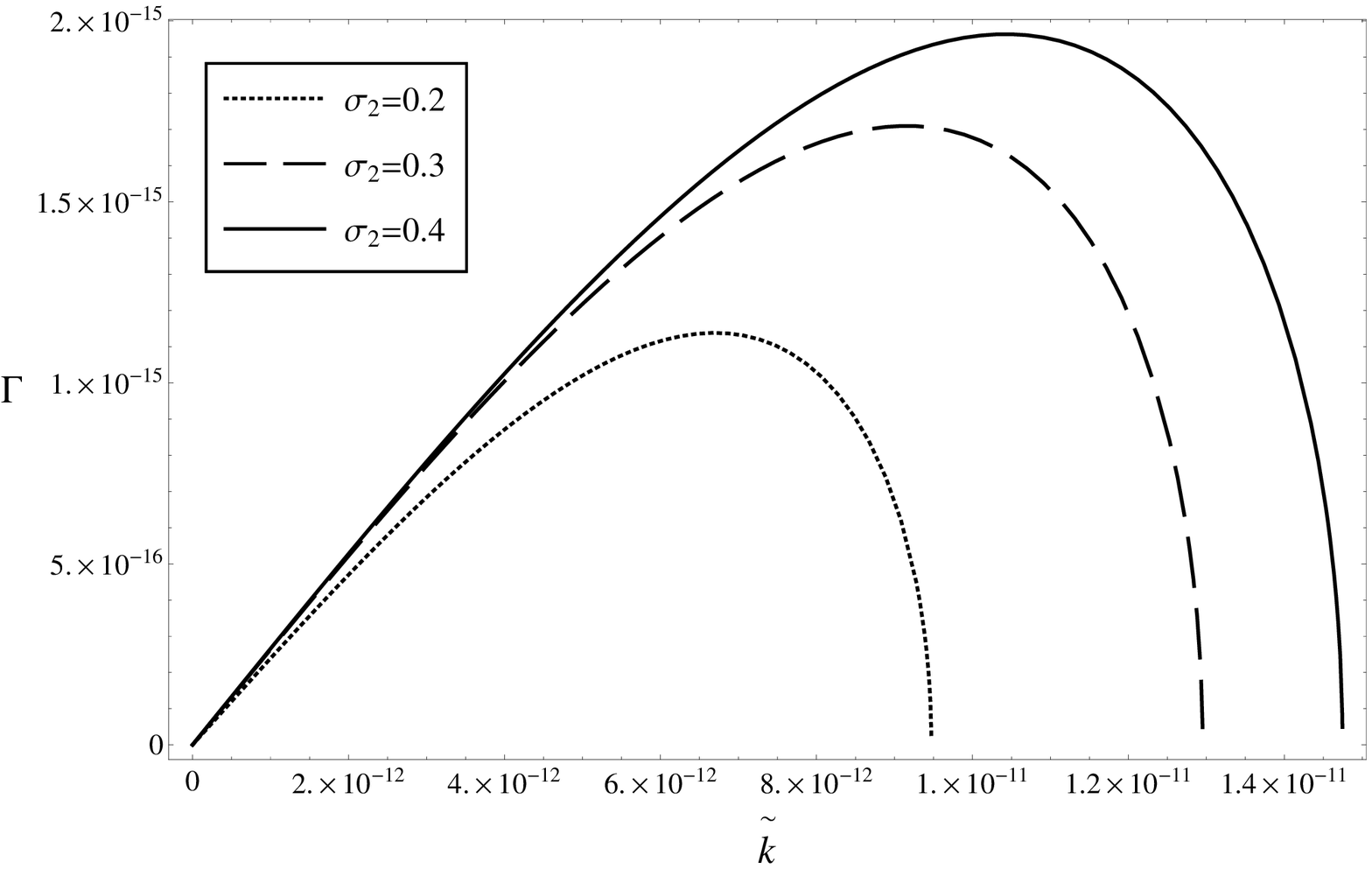}
  \caption{The variation of $\Gamma$ with $k$ for different values of $\sigma_2$;
  along with $\mu=1.5$, $\alpha = 3.67\times10^3$, $\gamma=2.16$, $\gamma/Z_l = 0.5$, $\sigma_1=3.68\times10^3$, $\lambda=0.2$, $k=0.0004$, and $\phi=0.8$.}
  \label{1Fig:F3}
  \vspace{0.8cm}
  \includegraphics[width=80mm]{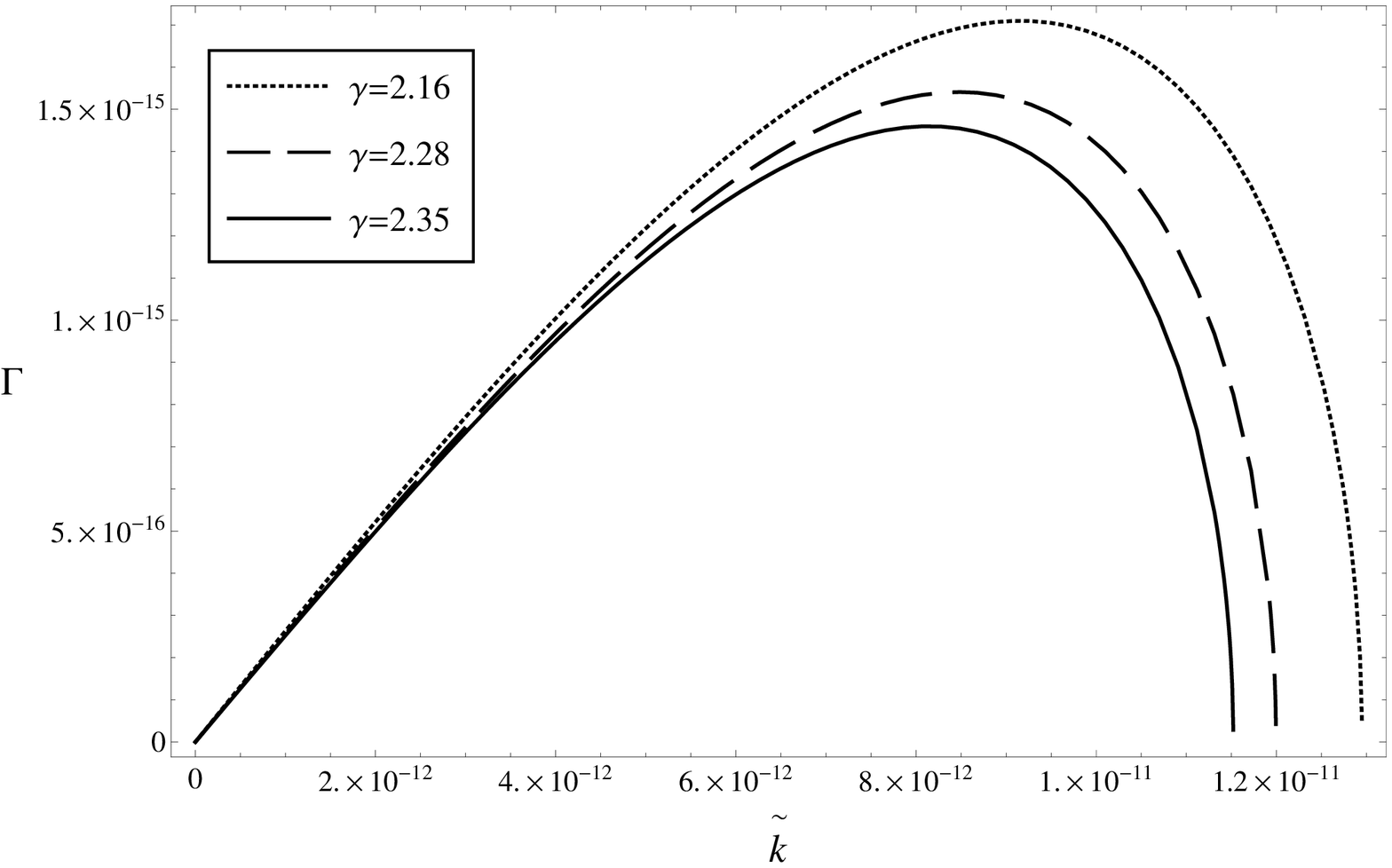}
  \caption{The variation of $\Gamma$ with $k$ for different values of $\gamma$;
  along with $\mu=1.5$, $\alpha = 3.67\times10^3$, $\gamma/Z_l = 0.5$, $\sigma_1=3.68\times10^3$, $\sigma_2=0.3$, $\lambda=0.2$, $k=0.0004$, and $\phi=0.8$.}
  \label{1Fig:F4}
\end{figure}
The effect of $\sigma_2$ and $\gamma$ on the growth rate are presented in Figs. \ref{1Fig:F3} and \ref{1Fig:F4},
where $\Gamma$ is plotted against $\tilde{k}$ and it is observed that (a) the growth rate ($\Gamma$) increases
with the increase in the value of positron number density $n_{p0}$, but decreases with increase of the  electron number density $n_{e0}$ (via $\sigma_2 = n_{p0}/ n_{e0}$);
(b) the maximum value of $\Gamma$ increases (decreases) with the decrease of $m_h$ ($m_l$) for the fixed value of $Z_l$ and $Z_h$ (via $\gamma = Z_l m_h/ Z_h m_l$);
(c) on the other hand, the maximum value of $\Gamma$ increases (decreases) with the decrease of $Z_l$ ($Z_h$) for the fixed value of $m_h$ and $m_l$ (via $\gamma = Z_l m_h/ Z_h m_l$).
So, the charge state and mass of the light and heavy ion plays an opposite role to manifest the $\Gamma$ in SG-DQPS.
The physics of this result is that the nonlinearity of the SG-DQPS increases (decreases) with the increase of the
value of $m_l$ or $Z_h$ ($m_h$ or $Z_l$) which enhance (suppress) the maximum value of the $\Gamma$.

The self-gravitating bright envelop solitons are generated in the modulationally unstable region (when $P/Q>0$) and
the solitonic solution of Eq. (\ref{1eq:29}) for the self-gravitating bright envelope solitons can be written as \cite{Chowdhury2018,Sultana2011,Schamel2002,Kourakis2005,Fedele2002}
\begin{eqnarray}
&&\hspace*{-0.0cm}\Phi(\xi,\tau)=\left[\psi_0~\mbox{sech}^2 \left(\frac{\xi-U\tau}{W}\right)\right]^{1/2}
\nonumber\\
&&\hspace*{1.0cm}\times \exp \left[\frac{i}{2P}\left\{U\xi+\left(\Omega_0-\frac{U^2}{2}\right)\tau \right\}\right],
\label{1eq:34}
\end{eqnarray}
\begin{figure}[t!]
  \centering
  \includegraphics[width=80mm]{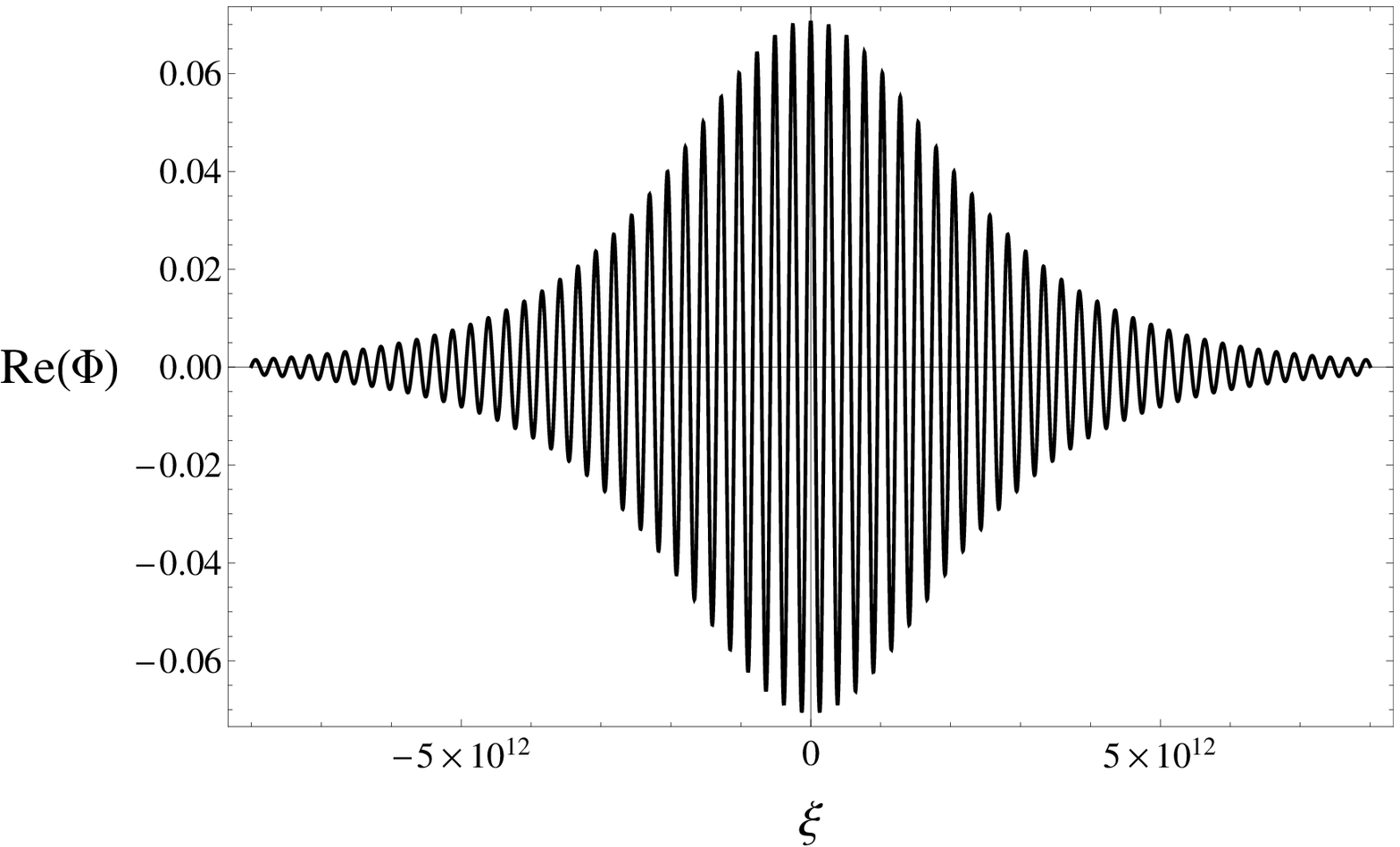}
  \caption{The variation of Re($\Phi$) with $\xi$ for bright envelope solitons;
  along with $\mu=1.5$, $\alpha = 3.67\times10^3$, $\gamma=2.16$, $\gamma/Z_l = 0.5$, $\sigma_1=3.68\times10^3$, $\sigma_2=0.3$, $\lambda=0.2$,
  $U=0.001$, $k=0.0004$, $\psi_0=0.005$, $\tau=0$, $\Omega_0=0.04$.}
  \label{1Fig:F5}
  \vspace{0.8cm}
  \includegraphics[width=80mm]{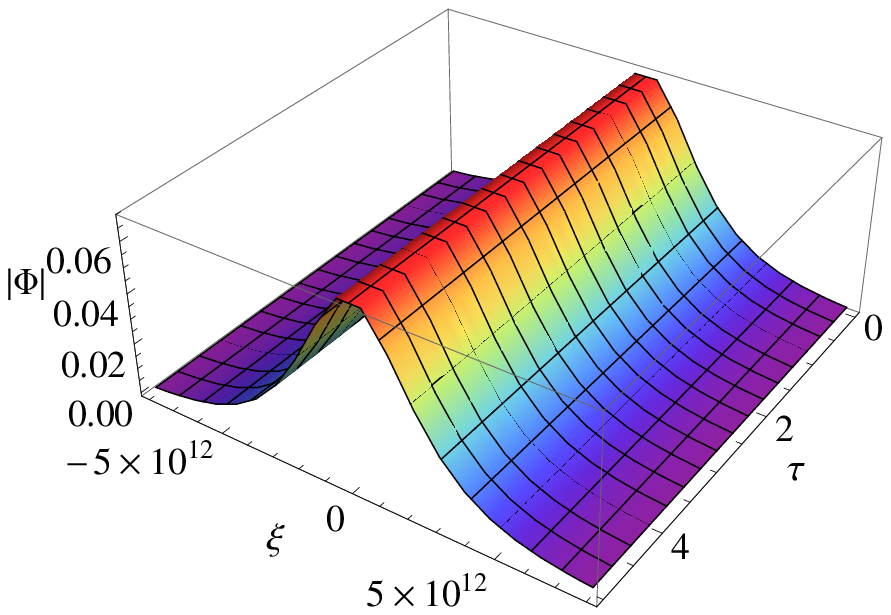}
  \caption{The variation of the $|\Phi|$ with $\xi$ and $\tau$ for bright envelope solitons;
  along with $\mu=1.5$, $\alpha = 3.67\times10^3$, $\gamma=2.16$, $\gamma/Z_l = 0.5$, $\sigma_1=3.68\times10^3$, $\sigma_2=0.3$, $\lambda=0.2$,
  $U=0.001$, $k=0.0004$, $\psi_0=0.005$, $\tau=0$, $\Omega_0=0.04$.}
  \label{1Fig:F6}
\end{figure}
where $U$ is the propagation speed of the localized pulse,
$W$ is the pulse width which can be written as $W=\sqrt{2|P/Q|/ \psi_{0}}$ ($\psi_0$ is the constant amplitude),
and $\Omega_0$ is the oscillating frequency for $U=0$. The self-gravitating bright envelop solitons which are
obtained from the numerical analysis of Eq. (\ref{1eq:34}), are depicted in Figs. \ref{1Fig:F5} and  \ref{1Fig:F6}. The bright envelop solitons remain
same as time($\tau$) passes, i.e., the self-gravitating bright envelop solitons are modulationally stable (please see Fig. \ref{1Fig:F6}).
\section{Discussion}
\label{1:Discussion}
In our above analysis, we have considered an unmagnetized realistic
laboratory or astrophysical SG-DQPS consisting of inertialess non-relativistically degenerate electron
and positron species, inertial non-relativistically degenerate light ion species as well as heavy ion species.
The NLS equation  has been derived by employing the well-known reductive perturbation method, which governs the evolution of nonlinear IAWs.
The notable informations that have been found from our theoretical investigation, can be pin-pointed as follows:
\begin{enumerate}
  \item{The angular wave frequency ($\omega$) of the IAWs exponentially decreases with the increase of $k$.
  On the other hand, the value of $\omega$ increases with the increase of $n_{e0}$ for the fixed value of $n_{l0}$ (via $\mu=n_{e0}/n_{l0}$).}
  \item{The IAWs will be modulationally stable (unstable) for that range of values of $k$ in which $P/Q$ is negative (positive), i.e., $P/Q<0$ ($P/Q>0$).}
  \item{The growth rate ($\Gamma$) increases with the increase in the value of positron number density $n_{p0}$, but decreases with
  increase of the  electron number density $n_{e0}$ (via $\sigma_2= n_{p0}/ n_{e0}$). On the other hand, the maximum value
  of $\Gamma$ increases (decreases) with the decrease of $m_h$ ($m_l$) for the fixed value of $Z_l$ and $Z_h$ (via $\gamma = Z_l m_h/ Z_h m_l$).
  Furthermore, the maximum value of $\Gamma$ increases (decreases) with the decrease of $Z_l$ ($Z_h$) for the fixed value of $m_h$ and $m_l$ (via $\gamma = Z_l m_h/ Z_h m_l$)}.
  \item{The self-gravitating bright envelop solitons remain same (modulationally stable) as time  passes.}
\end{enumerate}
The findings of this theoretical investigation may be useful for understanding the nonlinear structure (bright envelope solitons) of a SG-DQPS in space (viz.
neutron stars and white dwarf \cite{Chandrasekhar1931,Koester1990,Fowler1994,Koester2002,Zaman2017}).


\begin{thebibliography}{99}

\bibitem{Chandrasekhar1931} S. Chandrasekhar, Astrophys. J. \textbf{74}, 81 (1931)

\bibitem{Koester1990} D. Koester, G. Chanmugam, Rep. Prog. Phys. \textbf{53}, 837 (1990)

\bibitem{Fowler1994} R.H. Fowler, J. Astrophys. Astr. \textbf{15}, 115 (1994)

\bibitem{Koester2002} D. Koester, Astron. Astrophys. Rev. \textbf{11}, 33 (2002)

\bibitem{Zaman2017} D.M.S. Zaman, M. Amina, P.R. Dip,  A.A. Mamun 2017 Eur. Phys. J. Plus \textbf{132}, 457.

\bibitem{Atwater2007} H.A. Atwater, Sci. Am. \textbf{296}, 56 (2007)

\bibitem{Stockmann2011} M.I. Stockmann, Phys. Today \textbf{64}, 39 (2011)

\bibitem{Wolf2001} S.A. Wolf, D. Awschalom, R.A. Buhrman, Science \textbf{294}, 1488 (2001)

\bibitem{Manfredi2007} G. Manfredi, P.A. Hervieux, Appl. Phys. Lett. \textbf{91}, 061108 (2007)

\bibitem{Shapiro1983} S.L. Shapiro, S.A. Teukolsky, Black Holes, \textit{White Dwarfs and Neutron Stars: the
                         Physics of Compact Objects} (John Wiley \& Sons, New York, 1983)

\bibitem{Fletcher2006} R.S. Fletcher, X.L. Zhang,  S.L. Rolston, Phys. Rev. Lett. \textbf{96}, 105003 (2006)

\bibitem{Killian2006} T.C. Killian, Nature (London) \textbf{441}, 297 (2006)

\bibitem{Vanderburg2015} A. Vanderburg \textit{et al}., Nature (London) \textbf{526}, 546 (2015)

\bibitem{Witze2014} A. Witze, Nature \textbf{510}, 196 (2014)

\bibitem{Mamun2011} A.A. Mamun, P.K. Shukla, Europhys. Lett. \textbf{94}, 65002 (2011)

\bibitem{Asaduzzaman2017} M. Asaduzzaman, A. Mannan, A.A. Mamun, Phys. Plasmas \textbf{24}, 052102 (2017)

\bibitem{Mamun2017} A.A. Mamun, Phys. Plasmas \textbf{24}, 102306 (2017)

\bibitem{Chowdhury2018} N.A. Chowdhury, M.M. Hasan, A. Mannan, A.A. Manun, Vacuum  {\bf147}, 31 (2018)

\bibitem{Taniuti1969} T. Taniuti, N. Yajima, J. Math. Phys. \textbf{10}, 1369 (1969)

\bibitem{Chowdhury2017a} N.A. Chowdhury, A. Mannan, M.M. Hasan, A.A. Mamun,  Chaos \textbf{27}, (2017) 093105.

\bibitem{Sultana2011} S. Sultana, I. Kourakis, Plasma Phys. Control. Fusion {\bf53}, 045003 (2011)

\bibitem{Schamel2002} R. Fedele, H. Schamel, Eur. Phys. J. B {\bf27}, 313 (2002)

\bibitem{Kourakis2005} I. Kourakis, P.K. Shukla, Nonlinear Proc. Geophys. {\bf12}, 407 (2005)

\bibitem{Fedele2002} R. Fedele, H. Schamel, Eur. Phys. J. B \textbf{27}, 313 (2002)

\bibitem{Chowdhury2017b} N.A. Chowdhury, A. Mannan, A.A. Mamun,  Phys. Plasmas \textbf{24}, (2017) 113701.

\end{thebibliography}
\end{document}